\documentclass[10 pt, conference]{IEEEtran}
\IEEEoverridecommandlockouts

\usepackage{fancyhdr}

\usepackage[noadjust]{cite}

\usepackage{amsmath}
\usepackage{algpseudocode}
\usepackage{graphicx}
\usepackage{tabularx}
\usepackage{textcomp}
\usepackage{xcolor}
\usepackage{balance}
\usepackage{comment}
\def\BibTeX{{\rm B\kern-.05em{\sc i\kern-.025em b}\kern-.08em
    T\kern-.1667em\lower.7ex\hbox{E}\kern-.125emX}}
\usepackage[english]{babel}
\usepackage[utf8]{inputenc}
\usepackage{algorithm}
\usepackage{subcaption}
\usepackage{caption}
\usepackage{multirow}
\usepackage{graphics}
\usepackage{adjustbox}
\usepackage{soul}
\usepackage{enumitem}
\usepackage{tabularx}
\usepackage{array}
\usepackage{wrapfig}
\usepackage{url}
\usepackage{amsmath,amsfonts,amssymb}
\usepackage{graphicx}
\usepackage{pifont} % For tick and cross symbols
\usepackage{booktabs} % For better table formatting

\usepackage{booktabs}
\usepackage{graphicx}
\begin{document}

%\title{Mitigating the impact of hardware non-idealities for computing-in-memory for deeply scaled technologies}

\title{ReTern: Exploiting Natural \underline{Re}dundancy and Sign Transformations for Enhanced Fault Tolerance in Compute-in-Memory based \underline{Tern}ary LLMs}

%\title{Hardware Non-Idealities in Binary Neural Network Xbars and their Mitigation using Input-Weight Sparsification}

%\author{}
%\makeatletter
%\newcommand{\linebreakand}{%
%  \end{@IEEEauthorhalign}
%  \hfill\mbox{}\par
% \mbox{}\hfill\begin{@IEEEauthorhalign}
%}
%\makeatother

\author{\IEEEauthorblockN{Akul Malhotra and Sumeet Kumar Gupta}
\IEEEauthorblockA{
\textit{Purdue University}\\
West Lafayette, Indiana, USA \\
malhot23@purdue.edu}}

%\and
%\IEEEauthorblockN{Sumeet Gupta}
%\IEEEauthorblockA{
%\textit{Purdue University}\\
%West Lafayette, Indiana \\
%guptask@purdue.edu}

%}
%\end{comment}

%\author{\IEEEauthorblockN{Akul Malhotra and Sumeet Kumar Gupta}

% \author{\IEEEauthorblockN{}

% \and
% \IEEEauthorblockN{Ayush Arunachalam\IEEEauthorrefmark{1},
% Athulya Kizhakkayil\IEEEauthorrefmark{1}, Shamik Kundu\IEEEauthorrefmark{1}, Arnab Raha\IEEEauthorrefmark{2}}

% \and
% \IEEEauthorblockN{Suvadeep Banerjee\IEEEauthorrefmark{2},
% Robert Jin\IEEEauthorrefmark{3}, Fei Su\IEEEauthorrefmark{2} and Kanad Basu\IEEEauthorrefmark{1}}

%\IEEEauthorblockA{
% Department of Whatever,
% Whichever University\\
% Wherever\\
%\textit{Purdue University}, West Lafayette, IN}

%\thanks {{This work is supported by the Center for Brain-Inspired Computing (C-BRIC), one of six centers in JUMP, funded by Semiconductor
%Research Corporation. \textit{Corresponding Author}: \textit{Akul Malhotra}.} \newline %\hspace{5mm} Email: \textit{malhot23@purdue.edu}.
%}
%}

\maketitle
\thispagestyle{empty}
\pagestyle{empty}

\begin{abstract}
  Ternary large language models (LLMs), which utilize ternary precision weights and 8-bit activations, have demonstrated competitive performance while significantly reducing the high computational and memory requirements of full-precision LLMs. The energy efficiency and performance of Ternary LLMs can be further improved by deploying them on Ternary computing-in-memory (TCiM) accelerators, thereby alleviating the von-Neumann bottleneck. However, TCiM accelerators are prone to memory stuck-at faults (SAFs) leading to degradation in the model accuracy. This is particularly severe for LLMs due to their low weight sparsity.  To boost the SAF tolerance of TCiM accelerators, we propose ReTern that is based on (i) fault-aware sign transformations (FAST) and (ii) TCiM bit-cell reprogramming exploiting their natural redundancy. The key idea is to utilize FAST to minimize computations errors due to SAFs in +1/-1 weights, while the natural bit-cell redundancy is exploited to target SAFs in 0 weights (zero-fix). Our experiments on BitNet b1.58 700M and 3B ternary LLMs show that our technique furnishes significant fault tolerance, notably ~35\% reduction in perplexity on the Wikitext dataset in the presence of faults. These benefits come  at the cost of $<$ 3\%, $<$ 7\%, and $<$ 1\% energy, latency and area overheads respectively.  
\end{abstract}


\vspace{-0.1in}

\section{Introduction}
\label{sec:introduction}

The advent of large language models (LLMs) has revolutionized language-based applications, demonstrating state-of-the-art performance across various tasks. However, their resource-intensive nature poses significant challenges for efficient deployment on hardware, especially on resource-constrained edge devices \cite{llm_cost}. Moreover, as LLMs grow in size, the demands for storage and computation as well as the data movement between memory and processors increase, further exacerbating deployment challenges.

Computing-in-memory (CiM) has emerged as a popular hardware paradigm to alleviate the von-Neumann bottleneck that plague conventional AI accelerators. By performing in-memory vector-matrix multiplication (dot product) operation, the dominant kernel in LLM inference, CiM minimizes the data movement between the compute and memory units, providing energy and latency benefits. Various CiM-based LLM accelerators have been proposed, showcasing sizeable benefits over conventional von-Neumann based platforms such as GPUs \cite{cim_llm1,cim_llm2,cim_llm3,cim_llm4}.  

Another approach to enhance efficiency is quantization, which reduces model size by reducing the precision of the model parameters. Quantization techniques have been shown to achieve substantial reductions in storage needs, energy, and latency while maintaining accuracy \cite{llm_quant}. In fact, aggressive quantization schemes like binary/ternary precision quantization have also proven to achieve comparable model accuracy, while providing a massive increase in the energy efficiency, speed and storage density \cite{bitnet1,bitnet2,bitnet3}. Recently, a ternary LLM variant called Bitnet b1.58 has been proposed, that utilizes ternary precision weights and 8-bit activations \cite{bitnet2}. Experiments show that Bitnet b1.58 can match the accuracy of a full precision LLM (fp16) at 3B model size and beyond, making it a suitable candidate for resource-constrained applications.

By combining the benefits of CiM and ternary quantization, ternary CiM (TCiM) designs offer a compelling argument for LLM acceleration at the edge. TCiM accelerators feature massively parallel in-memory vector-matrix multiplications with ternary weights and/or activations and have been implemented using both SRAM and emerging non-volatile memory (NVM) technologies, demonstrating significant energy-latency-area advantages \cite{site_cim,tim_dnn,ternary1,ternary2}. While these TCiM accelerators have been evaluated on convolutional/recurrent neural networks (CNNs/RNNs), their benefits in performing massively parallel ternary in-memory vector-matrix multiplications can, in principle, be extended to ternary LLMs. 

However, despite these advantages, TCiM accelerators remain susceptible to manufacturing defects, particularly memory stuck-at faults (SAFs). SAFs irreversibly corrupt a portion of the memory bits, leading to the erroneous storage of model weights, which leads to inaccuracies in inference. The issue for SAFs are further worsened for emerging memory technologies, due to their relatively immature fabrication processes. Several solutions have been proposed to address this issue. However, most techniques require model re-training or fine-tuning \cite{training1,training2}, which can be prohibitively expensive for large models like LLMs \cite{ai_index}. Additionally, these approaches are often incompatible with CiM, not suitable for ternary precision, and introduce significant resource overhead due to the addition of large amounts of redundancy \cite{mapping1,mapping2,mapping3, redundancy3}. 

To address the limitations of previous solutions, we proposed TFix in a prior work \cite{tfix}. TFix is a training-free technique that mitigates the impact of memory faults for TCiM accelerators. TFix utilizes the natural redundancy present in TCiM bitcells as well as the weight sparsity in ternary precision DNNs to significantly reduce inference accuracy degradation due to SAFs. However, TFix has so far been evaluated only on ternary CNNs, and its effectiveness for other workloads, such as ternary LLMs, remains an open question.

In this work, we throw light onto the limitations of TFix on ternary LLMs deployed on TCiM accelerators and propose a new technique that counters these limitations. Our analysis of TFix reveals that it is less effective for ternary LLMs due to their inherently lower weight sparsity. To address this limitation, we propose ReTern, a novel approach that leverages fault-aware sign transformations (FAST) alongside the inherent redundancy of ternary bitcells. While TFix can only handle SAFs in zero weights, ReTern utilizes FAST to target +1/-1 weights and integrates FAST with TFix (zero-fix) to expand to coverage to all three weight types that can be affected by SAFs, In other words, the sign transformations in conjunction with zero-fix augment the ability of ReTern to minimize the impact of SAFs on the ternary LLM accuracy, while preserving the functionality of the vector matrix multiplications. Our key contributions include:

\begin{itemize}
    \item We analyze the effect of SAFs on two ternary LLMs: BitNet 1.58 (700M) and BitNet 1.58 (3B), deployed on TCiM hardware. 

    \item We assess the accuracy improvements offered by TFix and identify its limitations, particularly in ternary LLMs, in which weight sparsity is lower compared to convolutional ternary DNNs \cite{tfix,sparse_tnn,fat}. 

    \item We propose ReTern, which utilizes fault-aware sign transformations and the innate redundancy of ternary bit-cells to  improve the fault tolerance of ternary LLMs. We show that ReTern offers sizeable improvements in perplexity on the Wikitext dataset as well as accuracy on PIQA and ARC (easy) in the presence of SAFs.  

    \item We evaluate the hardware overheads of ReTern for SRAM, FeFET and ReRAM based TCiM accelerators, showing a minimal increase in energy, latency and area. 
\end{itemize}

%We observe that due to the reduced weight sparsity in Ternary LLMs (as compared to some convolutional ternary DNNs \cite{tfix,sparse_tnn,fat}), TFix has reduced effectiveness in enhancing the fault tolerance of ternary LLMs.

\section{Background and Related Works}
\label{sec:background}

In this section, we provide background on ternary LLMs and ternary CiM hardware. We then examine the impact of stuck-at faults (SAFs) on CiM-based DNN accelerators and review existing mitigation strategies along with their limitations. We also provide a brief overview of TFix and discuss its limitations, identifying key areas for improvement.

\subsection{Ternary Large Language Models (LLMs)}
\label{sec:ternary_llms}

 BitNet b1.58 is a ternary precision LLM in which the LLM weights are represented as signed ternary values: [-1,0,1]. Per-tensor weight quantization to ternary precision is performed by utilizing the \textit{absmean} quantization function: 

\begin{align}
    \tilde{W} &= \text{RoundClip}(\frac{W}{\gamma + \epsilon} , -1,1), \\
    \text{RoundClip}(x,a,b) &= \max(a, \min(b, \text{round}(x))), \\
    \gamma &= \frac{1}{nm} \sum_{ij} |W_{ij}|.
\end{align}
Where $\tilde{W}$ is the ternary weight, $W$ is the full precision weight, $\gamma$ is the mean absolute value of $W$ and $\epsilon$ is a small positive constant added to prevent division by zero.

Ternary LLMs are obtained using quantization-aware training (QAT), which is substantially costlier than post-training quantization (PTQ). Experiments in \cite{bitnet2} show that BitNet b1.58 can match the performance of the full precision (fp16) baseline from model size 3B and above. BitNet b1.58 utilizes 8-bit activations, although it is worth noting that works like \cite{bitnet3} are also exploring the possibility of reducing the activation bit precision from 8 to 4 bits. In this work, we evaluate the SAF tolerance of BitNet b1.58 with 700 million (BitNet b1.58 700M) and 3 billion ((BitNet b1.58 3B) parameters, when deployed on TCiM accelerators.

\subsection{Ternary Computing-in-memory (TCiM)}
\label{sec:ternary_cim}

\begin{figure}[t!]
\centering
  \includegraphics[width = 0.98\linewidth]{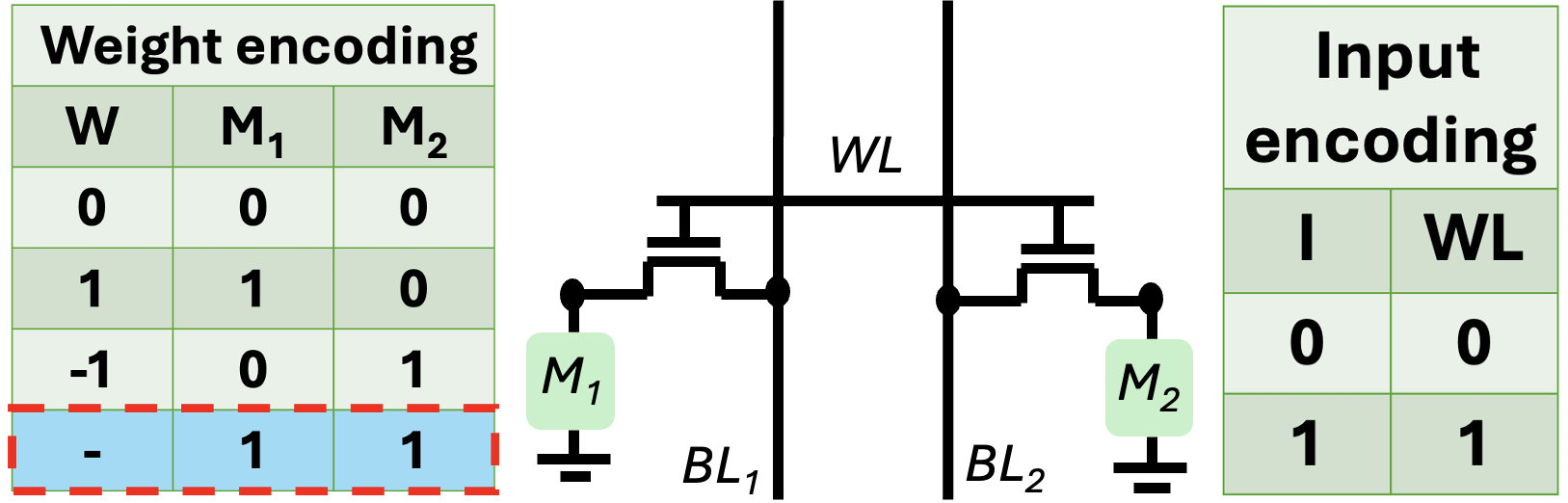}
  \caption{TCiM bitcell with its input and weight encoding. The boxed state (blue) in the weight encoding is an unused state (natural redundancy).}
  \label{fig:tcim_bitcell}
  %\vspace{-2.5mm}
\end{figure}

Various TCiM macros have been proposed using both CMOS and non-volatile memory (NVM) technologies, such as resistive RAM (ReRAM), magnetic RAM (MRAM), and ferroelectric FETs (FeFETs) \cite{site_cim,tim_dnn,ternary1,ternary2}. Broadly, TCiM macros can be categorized into two types: (1) signed ternary input with signed ternary weights, and (2) higher-precision input with signed ternary weights. For the former, two binary memory elements are utilized to realize a ternary weight (three states). For in-situ matrix-vector multiplications for signed inputs and weights, cross-coupling via access transistors can be used as proposed in \cite{site_cim}. For the latter design also, two binary memory elements are used per ternary bit-cell; however, cross-coupling is not required, as shown in Fig~\ref{fig:tcim_bitcell}. 

This work focuses on TCiM macros of type (2), aligning with the 8-bit activations of the ternary LLMs in our study. However, the SAF tolerance enhancement strategies proposed in this work are also applicable to TCiM macros based on design of type (1). Fig.~\ref{fig:tcim_bitcell} illustrates a common bitcell topology for signed ternary weights and higher precision inputs. Since a single ternary weight can have three possible states, two binary memory elements $M_1$ and $M_2$ are required. $M_1$ and $M_2$ could be CMOS or NVM-based. When $M_{1}$/$M_{2}$ is in '0' state, there is a high resistance path through it to ground, whereas '1' state provides a conducting/low resistance path. The ternary weights are stored using the differential weight encoding shown in Fig~\ref{fig:tcim_bitcell}. The high-precision activation $I$ is bit-streamed on $WL_{1}$ and is binary (0 or $V_{DD}$) per cycle. In case $I$ is signed, it can be bit-streamed in 2's complement encoding. Since two binary elements can represent four possible states, each bitcell has a natural redundancy in the form of an unused state, which corresponds to $M_{1} = M_{2} = 1$. We refer to this redundancy as natural since it is a consequence of implementing a ternary weight using two binary memory elements.

From Fig.~\ref{fig:tcim_bitcell}, we observe that the ternary weight ($W$) can be interpreted as $W = M_1 - M_2$, resulting in the scalar product $I \cdot W$ being the difference of two scalar products, $I \cdot M_1$ and $I \cdot M_2$. To obtain the scalar product of $W$ and $I$, the read bitlines $BL_1$ and $BL_2$ are precharged to the supply voltage $V_{DD}$. Then, the wordline $WL$ is asserted based on the value of $I$. A discharge of $\Delta$ on $BL_{1}$ ($BL_{2}$) signifies a scalar product of +1 (-1), and no discharge on $BL_{1}$ and $BL_{2}$ represents a scalar product of 0.  To compute the dot product $\sum_{i=1}^{L} I_i W_i$, multiple wordlines are activated simultaneously, and the cumulative voltage drops on $BL_1$ and $BL_2$ correspond to $x = \sum I_i M_{1_{i}}$ and $y = \sum I_i M_{2_{i}}$. Analog-to-digital converters (ADCs) digitize these values, and a digital subtractor computes the partial sum, $x - y$ \cite{tim_dnn}. The CiM outputs are then accumulated in accordance with their bit-significance.

In this work, we focus on TCiM macros designed with binary memory elements. It may be possible to design ternary bitcells with multilevel NVMs. However, since ternary weights are signed, such a design would require pre- and post-processing circuitry, such as adder trees, to realize the correct output. Additionally, multi-state memories have reduced distinguishability as compared to binary memories and thus, are more prone to CiM errors from variations and hardware non-idealities \cite{multistate}.

\subsection{Stuck-at faults (SAFs) in CiM-based DNN accelerators}
\label{sec:stuck_at_faults}

Aggressive scaling and the exploration of emerging memory technologies have made the study of stuck-at faults (SAFs) more critical than ever. Additionally, conventional fault mitigation strategies are not always directly applicable to CiM-enabled memories, necessitating exploration of CiM-compatible SAF tolerant techniques. SAFs cause binary memory elements to become permanently fixed at either ‘0’ (High-Resistance State, HRS) or ‘1’ (Low-Resistance State, LRS), resulting in stuck-at-0 (SA0) and stuck-at-1 (SA1) faults, respectively. SAFs can be classified as either masked or unmasked, depending on the stored data and the fault type. Unmasked faults occur when the stored value and the fault state differ (e.g., an SA1 fault in a location intended to store ‘0’), leading to errors. In contrast, masked faults occur when the stored value matches the fault state, and thus, are innocuous. It is evident that only the unmasked faults directly contribute to errors. 

Various solutions have been explored for mitigating the impact of SAFs in DNN accelerators. One approach is to enhance fault tolerance via training/finetuning the DNN. The work in \cite{training2} utilizes fault-aware retraining of the DNN weights to enhance accuracy, while the work in \cite{training1} incorporates drop-connect into the training phase to increase the resilience of DNNs to SAFs. Although both solutions are highly effective, DNN training is not only expensive, but also requires access to labeled training data, which may not always be available. 

A training-free strategy to tackle SAFs is to utilize fault-aware weight mapping. The weights are mapped onto the CiM macros in a way that minimizes the number of unmasked SAFs. The works in \cite{mapping1,mapping2,mapping3} show improved accuracies for inference with SAFs; however, these mapping strategies are not applicable to ternary weights. For example, the technique in \cite{mapping1} maps the most significant bits (MSBs) of the weights on to the fault-free locations while mapping the least significant bits (LSBs) on the faulty memory elements to reduce the impact of SAFs. However, this technique can not be applied to TCiM accelerators since ternary weights do not have MSBs and LSBs. 

\begin{figure}[t!]
\centering
  \includegraphics[width = 0.98\linewidth]{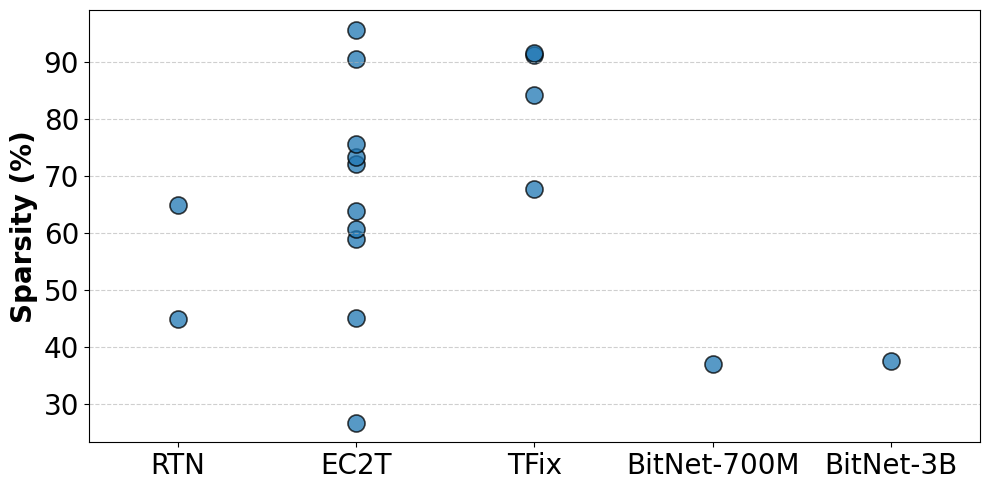}
  \caption{Ternary weight sparsities in DNNs, as reported in previous studies, compared to the sparsities observed in Bitnet 1.58B-700M and 3B ternary LLMs. Our findings indicate that ternary LLMs have relatively lower weight sparsities, making TFix less effective for these models.}
  \label{fig:ternary_sparsity}
  %\vspace{-2.5mm}
\end{figure}

Another training-free approach to improve SAF tolerance is the addition of artificial redundancy, where additional rows/columns are used to correct / compensate for the SAF-induced errors \cite{redundancy1,redundancy2}. However, such techniques may need sizeable overheads, with energy overheads ranging from $~24\%$ to $~112\%$ in \cite{redundancy1}. The approach in \cite{redundancy3} introduces an overhead-free solution by applying structured pruning to the DNN, freeing up space that is then used to add redundant rows and columns. However, this method necessitates model re-training to recover the accuracy lost due to pruning.

Our previous work TFix \cite{tfix} addressed the need for a training-free, low-overhead solution targeted for TCiM accelerators (more details in Section~\ref{sec:TFix}). However, the evaluation of TFix in \cite{tfix} is limited to convolutional neural networks (CNNs), for which TFix shows significant improvements in accuracy in the presence of SAFs. However, the question about the performance of TFix in LLMs is unaddressed. In this work, we evaluate the effectiveness of TFix on ternary LLMs deployed on TCiM accelerators, highlighting key limitations. Furthermore, we propose ReTern, that overcomes these limitations by incorporating fault-aware sign transformations. 

\begin{figure}[t!]
\centering
  \includegraphics[width = 0.98\linewidth]{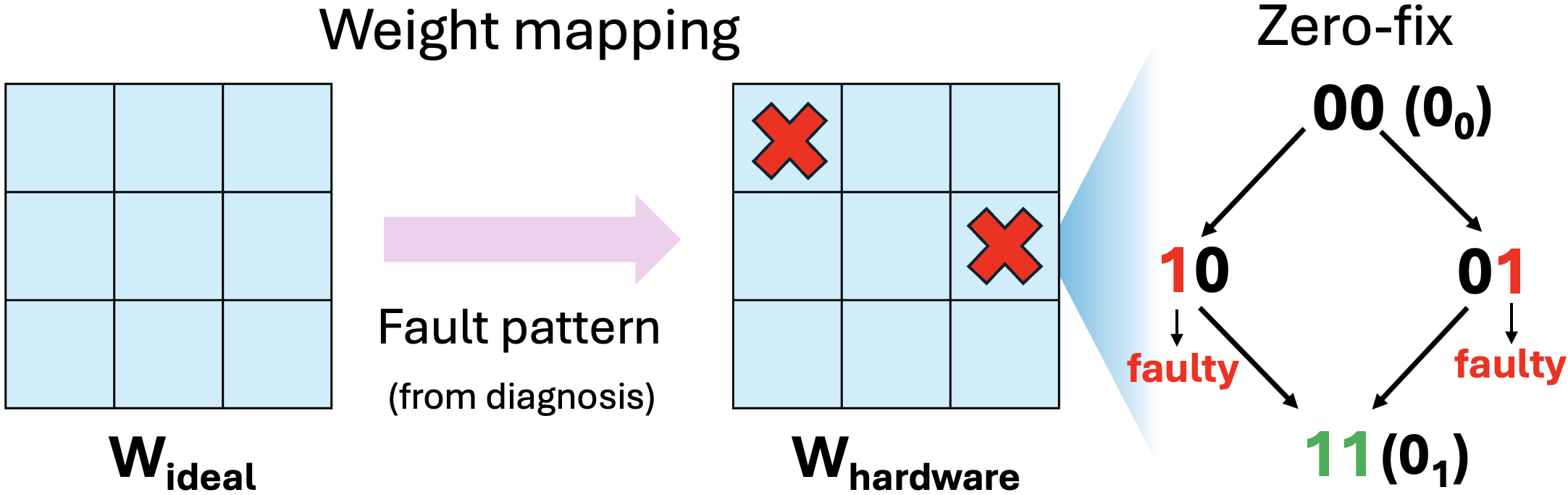}
  \caption{ Zero-fix, wherein, when an ideal zero weight ($0_0$), is erroneously stored as as $-1$ or $+1$ due to a SAF, it is fixed by reprogramming the memory cell to $0_1$.}
  \label{fig:tfix}
  %\vspace{-2.5mm}
\end{figure}

\subsection{TFix: SAF tolerant Solution for TCiM}
\label{sec:TFix}

TFix enhances SAF tolerance by exploiting the fact that a zero weight can be stored in two distinct ways, due to the inherent redundancy in TCiM bitcells (as discussed in Section~\ref{sec:ternary_cim}) \cite{tfix}. If a single SAF occurs in a TCiM bitcell storing zero, TFix can mitigate its impact by reprogramming the bitcell to the alternate state that also represents zero, effectively correcting the error (more details later).

TFix eradicates the impact of SAFs on zero weights without requiring any hardware overhead. However, TFix has some noteworthy limitations. Firstly, TFix can fix zero weights where only one memory element ($M_1$ or $M_2$) is impacted by an SAF. Fortunately, the probability of two SAFs in a single bitcell is very low, assuming a random distribution of SAFs. Another major limitation of TFix is that it can only correct SAF-induced errors in \textit{zero} weights. If an SAF occurs in a bitcell where $w_i$ is 1 or -1, the error cannot be fixed by Tfix due to the lack of alternative encoding options (like for the zero weight). Thus, the effectiveness of TFix is highly dependent on model weight sparsity. Specifically, the higher the percentage of zero weights, the more effective TFix becomes. As a result, SAF tolerance achieved by TFix becomes dependent on the attributes (e.g. weight sparsity) of specific ternary models. For example, previous works have reported high weight sparsity for various ternary precision models, with values as high as $\sim95\%$ (shown in Fig.~\ref{fig:ternary_sparsity}) \cite{rtn,sparse_tnn,tfix}, making them well suited for TFix. However, we observe that the Bitnet 1.58B ternary LLMs exhibit lower weight sparsities of 37.05\% and 37.55\% for the 700M and 3B parameter models, respectively (Fig.~\ref{fig:ternary_sparsity}). As a result, TFix would be relatively ineffective for BitNet 1.58B ternary LLMs, highlighting the need for an improved approach.
\section{Proposed solution: ReTern}
\label{sec:retern}

ReTern utilizes two techniques in conjunction to mitigate SAF impact: zero-fix and fault-aware sign transformation (FAST). Zero-fix, which is essentially TFix, corrects SAF-induced error in zero weights by leveraging the natural redundancy in TCiM bitcells. (Note, in this paper, we use the name zero-fix instead of TFix to avoid confusion with the previous TFix approach which explicitly relies on weight sparsity). The key operation that makes ReTern effective, irrespective of the sparsity of the network, is FAST. FAST minimizes the unmasked faults in -1/1 weights by selectively applying sign flips to weight columns in TCiM arrays, while maintaining the correct MVM functionality. Both zero-fix and FAST are complementary and do not interfere with each other, since zero-fix and FAST target zero and non-zero weights, respectively. 

\begin{figure}[t!]
\centering
  \includegraphics[width = 0.98\linewidth]{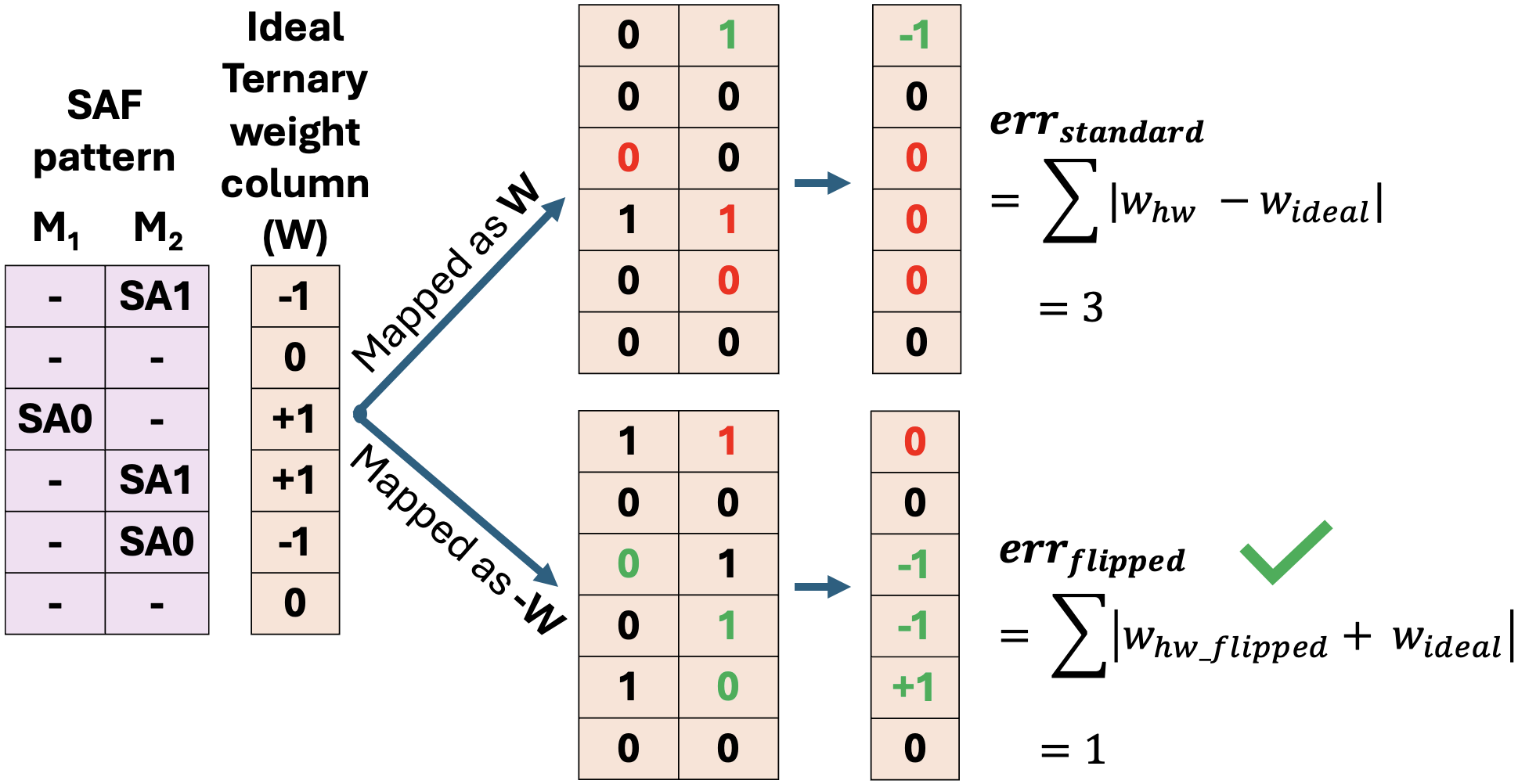}
  \caption{Fault-aware sign transformations being applied to a TCiM weight column.}
  \label{fig:weight_transformation}
  %\vspace{-2.5mm}
\end{figure}

\subsection{Zero-Fix}

\label{sec:zero_fix}

As discussed in Section~\ref{sec:ternary_cim}, TCiM bitcells have a natural redundancy in the form of an unused state, corresponding to $M_{1} = M_{2} = 1$. We observe that if the 11' state were to be used for scalar product computation, it would functionally mimic the behavior of the 00' (0 weight) state \cite{tfix}. This is because the '11' state would produce a voltage drop of $\Delta$ on both $BL_1$ and $BL_2$ for $I=1$. Since the scalar product is represented by the difference of the voltage drops in $BL_1$ and $BL_2$, the equal drops produced by the '11' state on $BL_1$ and $BL_2$ would cancel out, leading to an output of 0. Note, '11' represents the zero weight for both $I=0$ (no drop on both $BL_1$ and $BL_2$ since the access transistors are off) and $I=1$ (as discussed above). Thus, there are two ways to store the '0' weight, $M_{1} = M_{2} = 0$ ('00') or $M_{1} = M_{2} = 1$ ('11'), which we refer to as $0_0$ and $0_1$, respectively.

Zero-fix utilizes this natural redundancy to correct the errors due to SAFs in the zero weights. If a zero weight is incorrectly stored due to a single SAF (Fig.~\ref{fig:tfix}) in the bitcell, we utilize the fault-free memory element in the bitcell to eliminate the error due to the SAF. For example, if $w_{hw}$ (intended to be stored as $0_0$ ($w_{i} = 0$)), is in the ’01’/’10’ state due to a stuck-at 1 fault in $M_2$/$M_1$, it can be transformed to $0_1$ by programming $M_1$/$M_2$ to ’1’. Thus, the impact of the hard fault on the TCiM computation would be eliminated by rewriting the faulty zero weights to $0_1$.

\subsection{Fault-aware sign transformation (FAST)} 

While Zero-Fix effectively handles errors in zero weights, FAST mitigates SAF-induced errors in +1 and -1 weights, which make up the majority of ternary LLM weights. This enhancement overcomes the key limitation of TFix.

Fig.~\ref{fig:weight_transformation} illustrates how the sign transformation operation is applied to a single column of the TCiM memory array. After diagnosing the fault pattern of the column (discussed in the next sub-section), we choose whether to store the weights as they are ($W$), or the negative of all the weights in the column ($-W$). We refer to these two as the standard and the flipped options, since multiplication by-1 is equivalent to flipping the sign of the non-zero weights (+1 to -1 and vice-versa). We make this decision based on which option yields larger number of \textit{masked} faults (discussed in Section~\ref{sec:stuck_at_faults}) and thus fewer erroneous perturbations due to SAFs.  In the example illustrated in Fig.~\ref{fig:weight_transformation}, the flipped option is more suitable to store the weight column given the fault pattern. To obtain the correct vector-matrix multiplication with the flipped option, the TCiM output must be multiplied by -1, since $\sum_{i=1}^{L} I_i (-W_i) = -1*\sum_{i=1}^{L} I_i W_i$. Thus, it is essential to keep track of which columns in the TCiM memory array are flipped, for which we introduce a vector $col\_flip$. For a TCiM array with $m$ columns, the size of $col\_flip$ is $m$ bits. If the $i^{th}$ column is stored in the flipped (standard) form, the value of $col\_flip[i]$ is 1 (0). 

\subsection{ReTern algorithm}

\begin{algorithm}[t]  % 't' tries to place the algorithm at the top of the column
    \caption{ReTern Algorithm}
    \begin{algorithmic}[1]
        \State \textbf{Input:} Ideal TCiM array weight $W_{ideal}$
        
        \State \textbf{Output:} Optimized weight mapping $W_{mapped}$

        \State \textbf{Step 1:} Obtain SAF information from diagnosis.

        \State \textbf{Step 2:} Calculate $W_{hardware}$ from SAF information and $W_{ideal}$

        \State \textbf{Step 3:} Fault-Aware Sign Transformation (FAST)

        \For{each column in each TCiM memory array}
        
            \State \(err_{standard} \gets \sum |w_{hw} - w_{ideal}|\)
            
            \State \(err_{flipped} \gets \sum |w_{hw\_flipped} + w_{ideal}|\)

            \If{\(err_{flipped} < err_{standard}\)}
            
                \State \(W_{mapped-column} \gets -\,W_{ideal-column}\)
                
                \State \(col\_flip \gets 1\)
            \Else
            
                \State \(W_{mapped-column} \gets W_{ideal-column}\)
                
                \State \(col\_flip \gets 0\)
                
            \EndIf
        \EndFor

        \State \textbf{Step 4:} Zero-Fix

        \For{each weight \(w\) in \(W_{ideal}\) where \(w=0\)}
        
            \If{\(w_{hardware} \neq 0\)}
            
                \State Store \(w_{mapped} = 0_{1}\)
                
            \Else
            
                \State Store \(w_{mapped} = 0_{0}\)
                
            \EndIf
        \EndFor

        \State \textbf{Return} \(W_{mapped}\)
    \end{algorithmic}
    \label{algo:retern}
\end{algorithm}

The ReTern algorithm consists of four stages: weight mapping, diagnosis, FAST and zero-fix. In the weight-mapping step, we partition the ideal ternary weight matrix $W_{ideal}$ and map the submatrices onto multiple TCiM memory arrays. In the fault diagnosis stage, we utilize fault testing techniques to determine the position and type of SAFs in the TCiM arrays \cite{fault_testing}. Using the SAF information and $W_{ideal}$, we can determine the hardware-mapped weights $W_{hardware}$, which may contain incorrect weight values due to SAFs. 

In the FAST stage, we first calculate the SAF-induced error for each \textit{column} of each TCiM memory array. This error is measured as:

\begin{equation}
    err_{standard} = \sum|w_{hw} - w_{ideal}|
\end{equation}

where $w_{hw}$ represents the hardware-mapped weights, and $w_{ideal}$ represents the ideal weights. We then compute the SAF-induced error for the flipped weight configuration:

\begin{multline}
   err_{flipped} = \sum|(-w_{hw\_flipped}) - w_{ideal})| \\
   = \sum|w_{hw\_flipped} + w_{ideal}|
\end{multline}

Next, we compare $err_{standard}$ and $err_{flipped}$, selecting the mapping option that yields a smaller error. Concurrently, we set or reset the corresponding column flip ($col\_flip$) for each TCiM array based on whether the flipped or the standard column is used. 

Finally, we perform the zero-fix step and determine for each zero weight whether it should be stored in the $0_0$ or $0_1$ form, based on the individual SAF information (as discussed in Section~\ref{sec:zero_fix}). It should be noted that the FAST and zero-fix steps effect a \textit{mutually exclusive} set of weights. Zero-fix is designed to mitigate the impact of SAFs in zero weights, while FAST targets non-zero weights. This is because the transformations utilize sign-flipping, which does not effect zero weights, since $-1*0 = 0$. Thus, by utilizing FAST and zero-fix in conjunction in ReTern, we address the limitations of TFix without introducing any new complications. Additionally, due to their mutually exclusive nature, FAST and zero-fix steps can be performed in any order. ReTern is summarized in Algorithm~\ref{algo:retern}. 

In the next subsection, we discuss the hardware modifications needed for the TCiM array to support ReTern. It is important to note that the execution of the ReTern algorithm, which determines the optimal weight mapping, does not add to the inference overhead of the TCiM accelerator. This is because the algorithm execution is performed offline, prior to programming the weights into the TCiM array.

\begin{figure}[t!]
\centering
  \includegraphics[width = 0.98\linewidth]{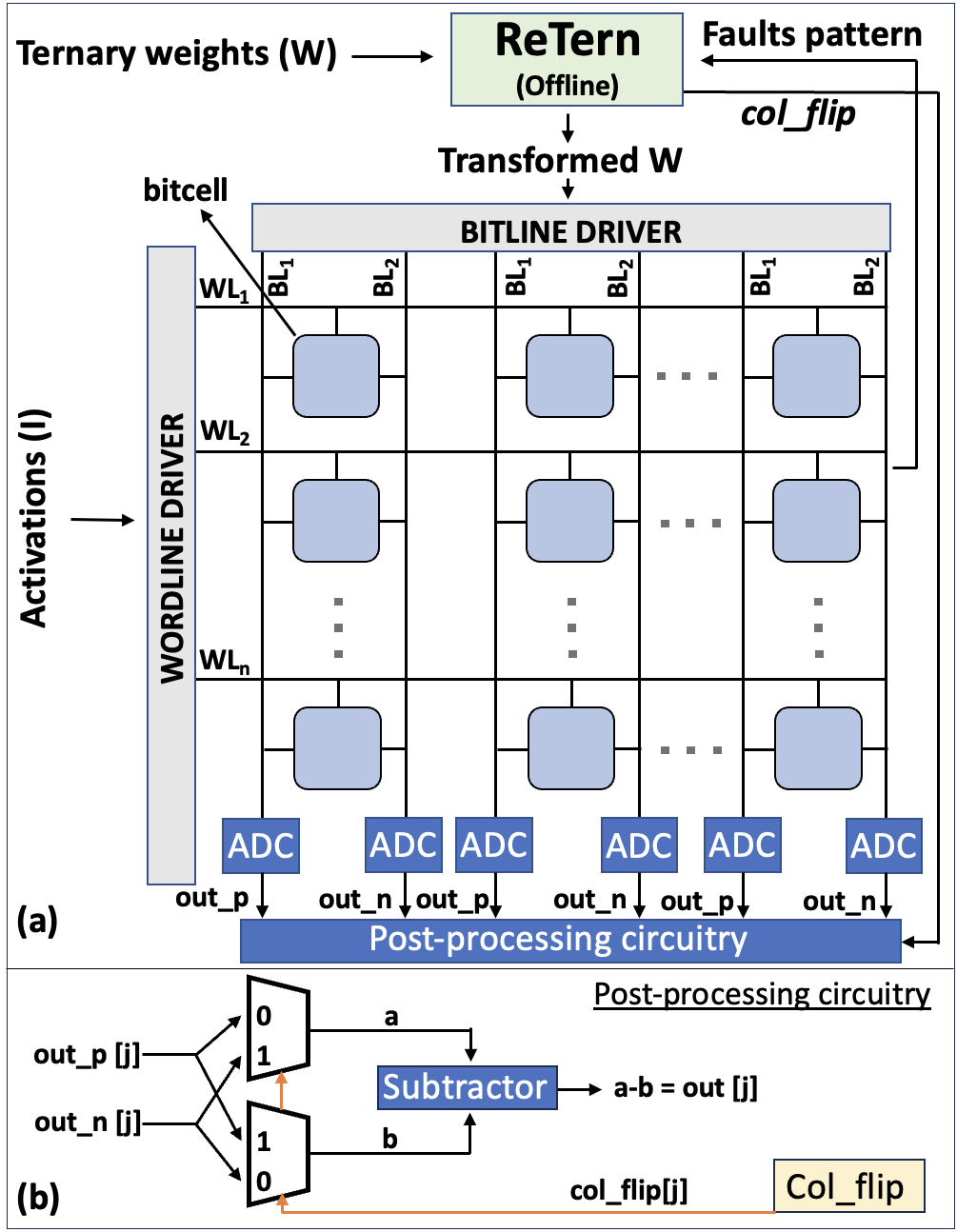}
  \caption{(a) Modified TCiM array to make it compatible with ReTern. (b) shows the post-processing required to obtain the correct dot-product. Two 2:1 multiplexers are added to datapath to choose whether the output or the negation of the output is to be computed, based on the value of the $col\_flip$ bit for the column.}
  \label{fig:tfix_hardware}
  %\vspace{-2.5mm}
\end{figure}

\subsection{Hardware Implementation}
\label{sec:hardware_imp}

Once ReTern is applied on the pre-trained ternary weights, the resultant weights are programmed onto the TCiM accelerator for deployment. Fig.~\ref{fig:tfix_hardware} (a) illustrates a TCiM memory array modified to support ReTern. To enable compatibility, additional peripheral circuitry is required. First, a near-memory register is required to store the $col\_flip$ vector. Second, peripheral circuitry is required for post-processing the output. Recall from Section~\ref{sec:ternary_cim}, the final dot product $x-y$ is computed using a subtractor, that subtracts the ADC outputs from the two bitlines. We use two 2:1 multiplexers to choose whether to compute $x-y$ or $y-x$, based on the value of $col\_flip$ value for that column(Fig.~\ref{fig:tfix_hardware}(b)). This is effectively the same as multiplying the final output with -1 (1) if the $col\_flip$ bit is 1 (0). (Recall, to maintian the correct MVM functionality, a column flip must be followed by negating the output).  Thus, the overheads of ReTern are minimal, requiring just one extra register for $col\_flip$, and two 2:1 multiplexers to post-process the outputs (details in Section~\ref{sec:results}). Note that the subtractor is not an overhead, since it is required to compute the dot product even in the baseline design \cite{site_cim,tim_dnn}. Note that the additional hardware required by ReTern is due to FAST, while zero-fix does not need any hardware modifications.

Both the efficacy and overheads of ReTern depend on the TCiM array size. Since the fault-aware sign transformations are carried out at the column granularity, the more the number of rows in the array, the less effective the transformations will be. However, for an $n$x$m$ size TCiM array, the memory overhead associated with $col\_flip$, which is of length $m$, is $\frac{m}{n*m} = \frac{1}{n}$. Thus, there exists a trade-off between ReTern performance and overhead with respect to the number of rows in the TCiM array. We aim to quantitatively explore this trade-off in a future work. In this work, we analyze the efficacy of our technique and the hardware overhead for a fixed array size of 64 x 64. It is important to note that this trade-off only exists for the sign transformation step of ReTern. The zero-fix step operates at the per-weight granularity, and is not directly effected by the array size.  

\section{Results}
\label{sec:results}

\subsection{Evaluation Framework}

In this section, we quantitatively investigate the SAF tolerance provided by ReTern for ternary LLMs. We conduct our experiments on the BitNet 1.58b 700M and 3B models, and we obtain the pretrained weights from Hugging Face \cite{700m_model,3b_model}. We evaluate our models on three tasks: Wikitext, PIQA \cite{piqa}, and ARC (easy) \cite{arc}. We compare the perplexity for Wikitext in the presence of SAFs for the baseline design (no fault tolerance), design using TFix \cite{tfix} (akin to using only zero-fix in ReTern), design with only FAST and the proposed ReTern (FAST + zero-fix). Similarly, for PIQA and ARC (easy), we compare the accuracies.  We select PIQA and ARC (easy) over other tasks for our evaluation because the ternary models exhibit reasonable software (ideal) accuracies on them. In contrast, other tasks, such as ARC (challenge), ternary LLMs yield software accuracies close to random guessing, making it difficult to clearly assess the impact of SAFs. (Note that the random-guess accuracies on PIQA and ARC (easy) are 50\% and 25\%, respectively). As discussed in Section~\ref{sec:hardware_imp}, we utilize multiple TCiM arrays of size 64x64 to map the model weights. The SAFs are injected randomly and uniformly across each of the TCiM arrays, as in similar works \cite{redundancy1}. 20 Monte Carlo fault injection experiments are performed for two SAF rates: 5\% and 10\%.  

An important noteworthy point about LLMs is that they perform two types of matrix multiplications during inference: (i) in the feedforward layers and (ii) in the self-attention layers. In the feedforward layers, matrix multiplication occurs between static weight matrices and dynamic input activations. In contrast, the self-attention layers involve matrix multiplications between dynamically generated query, key, and value matrices, which are computed on-the-fly based on the input representations. Given that non-volatile memories such as ReRAMs, FeFETs etc. have limitations such as large write latency/energy and limited endurance, TCiM arrays designed with them may not be suitable for self-attention. Thus, works like \cite{heterogenous} utilize \textit{digital} compute cores for the self-attention layers, while using CiM-based memory arrays for the feedforward layers. We also adopt such an architecture for our TCiM accelerator design, thus considering the self-attention layers to be SAF-free, with all the SAFs occurring in the feedforward layer ternary weights.  

\begin{figure}[t!]
    \centering
    \begin{subfigure}[b]{\linewidth}
        \centering
        \includegraphics[width=0.98\linewidth]{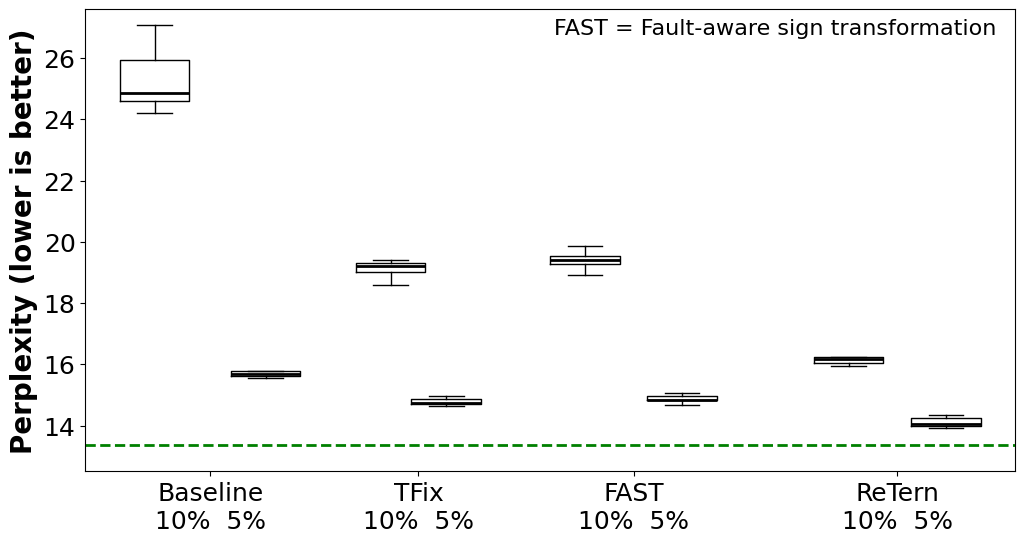}
        \caption{Wikitext}
    \end{subfigure}
    \hfill
    \begin{subfigure}[b]{\linewidth}
        \centering
        \includegraphics[width=0.98\linewidth]{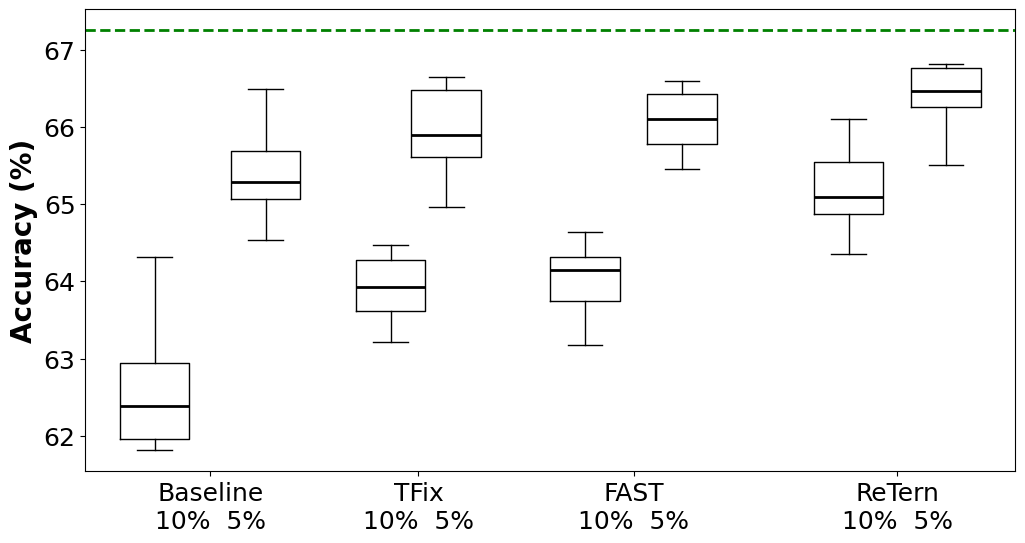}
        \caption{PIQA}
    \end{subfigure}
    \hfill
    \begin{subfigure}[b]{\linewidth}
        \centering
        \includegraphics[width=0.98\linewidth]{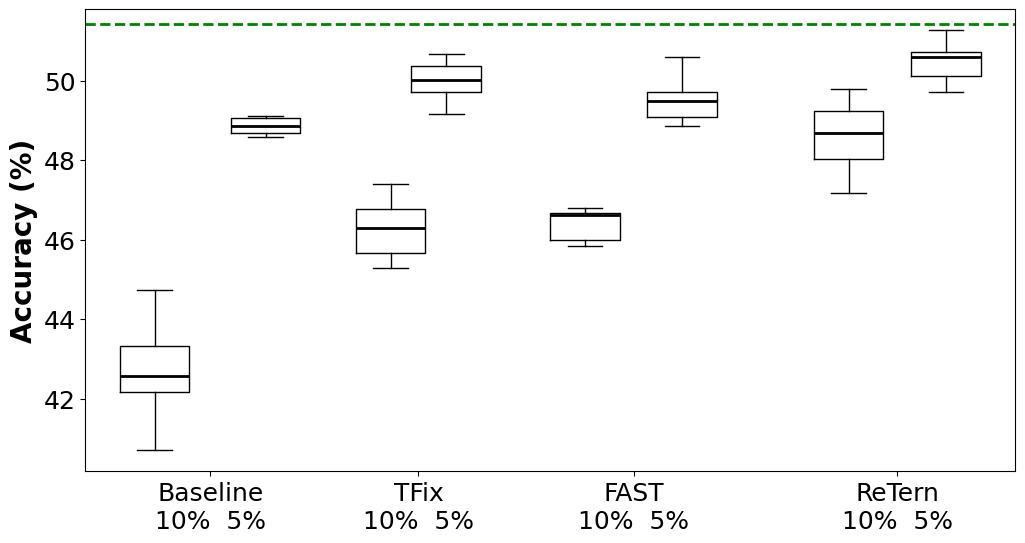}
        \caption{ARC (easy)}
    \end{subfigure}
    \caption{\centering Performance results for the 700M BitNet 1.58b model. SAF rates of 5\% and 10\% are evaluated.}
    \label{fig:700m_results}
\end{figure}

\begin{figure}[t!]
    \centering
    \begin{subfigure}[b]{\linewidth}
        \centering
        \includegraphics[width=0.98\linewidth]{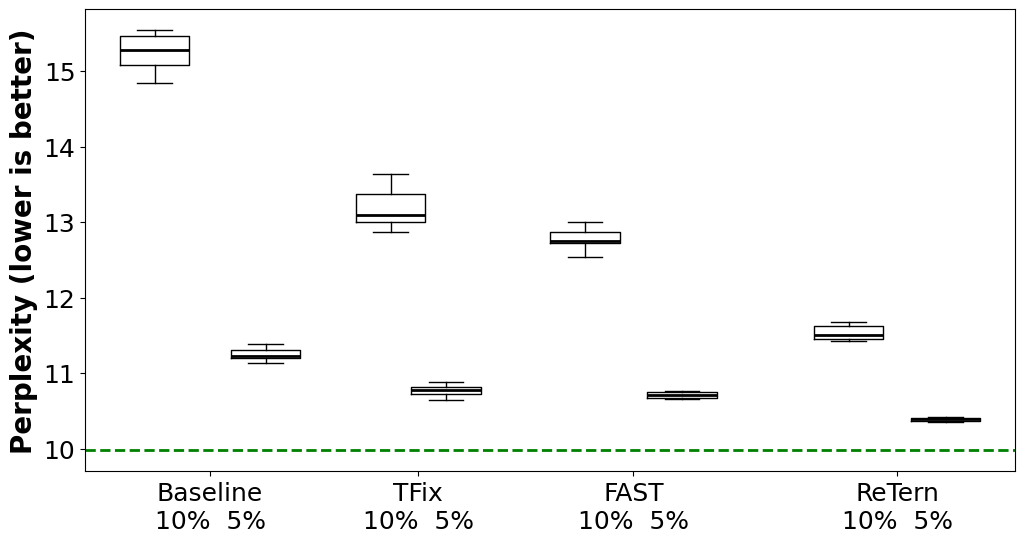}
        \caption{Wikitext}
    \end{subfigure}
    \hfill
    \begin{subfigure}[b]{\linewidth}
        \centering
        \includegraphics[width=0.98\linewidth]{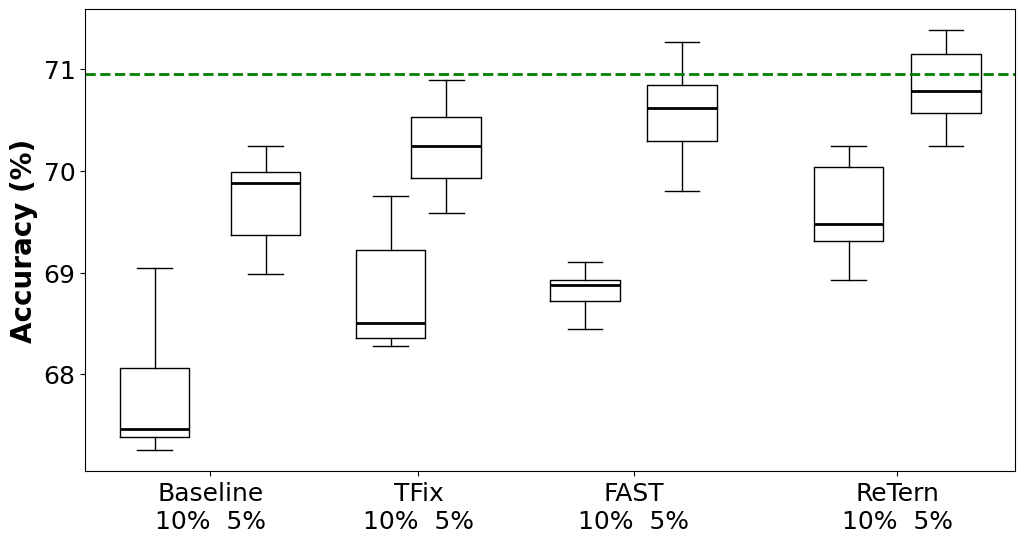}
        \caption{PIQA}
    \end{subfigure}
    \hfill
    \begin{subfigure}[b]{\linewidth}
        \centering
        \includegraphics[width=0.98\linewidth]{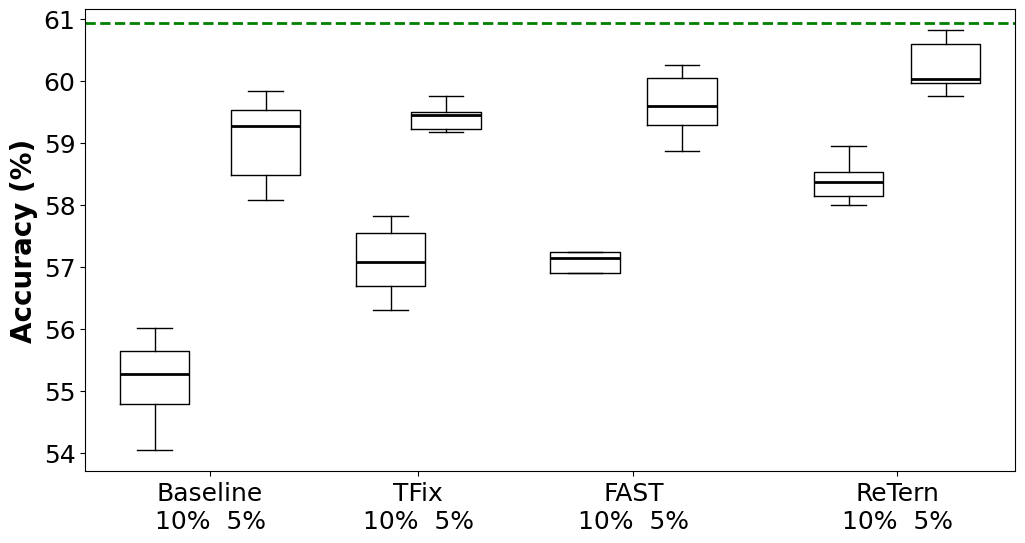}
        \caption{ARC (easy)}
    \end{subfigure}
    \caption{\centering Performance results for the 3B BitNet 1.58b model. SAF rates of 5\% and 10\% are evaluated.}
    \label{fig:3b_results}
\end{figure}

\subsection{SAF Tolerance}

Fig.\ref{fig:700m_results} and Fig.\ref{fig:3b_results} present box-and-whisker plots illustrating the impact of Stuck-at Faults (SAFs) on the BitNet 1.58b 700M and 3B models, respectively. The green dashed lines in each figure indicate the software/ideal accuracy and perplexity. Let us start with the baseline model. At an SAF rate of 10\% and for the 700M model, we observe an increase in Wikitext perplexity from approximately 12 to 26 and accuracy degradation of around 4\% and 8\% on PIQA and ARC (easy), respectively. In contrast, the larger 3B model shows a more moderate increase in Wikitext perplexity (from about 10 to 15) and accuracy degradations of approximately 3\% on PIQA and 6\% on ARC (easy)) at the same 10\% SAF rate. Thus, we observe that the larger 3B parameter model exhibits greater robustness to SAFs compared to the 700M parameter model. This may be attributed to the larger redundancy present in the bigger model. 

Next, we observe that both TFix and FAST have a similar effectiveness in mitigating the impact of SAFs. At the SAF rate of 10\% (5\%), both techniques individually improve the Wikitext perplexity by around 23\% (6\% ) on average for the 700M model and by 15\% (5\%) for the 3B model. For the PIQA dataset, TFix and FAST improve accuracy by approximately 2\% and 1\%, respectively, for the 700M and 3B models at 10\% SAF rate. On the ARC (Easy) dataset, these techniques individually yield an average accuracy improvement of $\sim$4\% for the 700M model and $\sim$2\% for the 3B model at 10\% SAF rate.

Finally, we observe that ReTern enhances the fault tolerance  through the integration of FAST and zero-fix. For the SAF rate of 10\% (5\%), ReTern furnishes an average Wikitext perplexity reduction of around 35\% (10\% ) on average for the 700M model and of 25\% (8\%) for the 3B model. Further, ReTern improves accuracy of PIQA by around 4\% and 2\%, and of ARC (easy) by 6\% and 5\%, for the 700M and 3B models, respectively, at a 10\% SAF rate. These results demonstrate that ReTern effectively combines the advantages of zero-fix and FAST, delivering the most significant accuracy improvements in the presence of SAFs for various ternary LLMs.

\subsection{Hardware Overhead}

We evaluate the hardware implications of ReTern at the memory macro level (TCiM array + peripheral circuits). The overall system-level impact may be lower when considering additional components beyond the memory macro. In line with our accuracy analysis, we design 64×64 TCiM arrays based on three memory bitcell technologies: 8T-SRAM, 1T-1ReRAM, and 1FeFET. The TCiM arrays employ current-sensing to generate array outputs. For the 8T-SRAM array design, we utilize the 7nm Predictive Technology Model (PTM) \cite{ptm}. For the ReRAM, we employ an experimentally calibrated compact model of Al-doped $HfO_{X}$ ReRAM from \cite{rram}. For the FeFET, we utilize a compact model of $Hf_{0.5}Zr_{0.5}O_{2}$-based FeFET
in which the ferroelectric layer is modeled with modified Preisach equations \cite{fefet} and is coupled with a 7nm transistor. This model is calibrated with experiments in \cite{fefet_experiments} and validated using self-consistent phase-field simulations \cite{fefet_sims}. To accurately estimate the energy, latency, and area of these arrays, we first develop custom layouts following the methodology presented in \cite{bitcells}. Once the array dimensions are determined in terms of gate, metal, and fin pitches, we estimate parasitic wire resistances \cite{para_res} and capacitances \cite{para_cap}. Subsequently, these estimations are incorporated into SPICE simulations to determine the energy, latency, and area (derived directly from custom layout dimensions) of the arrays. 

To estimate the energy, latency, and area of the TCiM array peripherals—including the row decoder,flash ADC, and subtractor—as well as the additional multiplexers (MUXes) and register required for ReTern, we use NeuroSim \cite{neurosim}. We again utilize the 7nm predictive technology model (PTM) for the peripherals and the post-processing circuitry \cite{ptm}.

To keep other TCiM array non-idealities like the effect of parasitic resistances and device non-linearities in check, we utilize partial word-line activation (PWA), simultaneously activating 16 rows (out of 64) at a time. PWA also reduces the required ADC precision, enabling us to use 4-bit flash ADCs, albeit at the cost of latency. To further minimize ADC area overhead, we share one set of column peripherals (ADCs + post-processing circuitry) across eight columns, again trading off latency for improved area efficiency.

Table~\ref{table:overhead} presents the overheads of ReTern relative to the baseline (i.e. with no fault tolerant technique). As discussed in Section~\ref{sec:hardware_imp}, the primary source of overhead stems from FAST. Specifically, the additional hardware introduced by FAST in ReTern (illustrated in Fig.~\ref{fig:tfix_hardware}) consists of two 2:1 multiplexers per column to route the two TCiM array outputs into the subtractor and a single register per TCiM array to store the $col\_flip$ bit (an overhead of 1 bit per column). Our evaluation shows that ReTern incurs a modest energy overhead of 2.0\%-2.2\% and a latency increase of 3.2\%-6.6\% compared to the baseline. These overheads remain negligible in the broader context of CiM energy and latency, which are primarily dominated by ADCs and bitline switching. Furthermore, we estimate the area overhead of ReTern to be minimal, remaining below 1\% across all three memory technologies. This is again attributed to the dominant area contributions from ADCs and memory arrays. Overall, ReTern offers significant SAF tolerance with only minor resource overhead, making it a practical enhancement for ternary LLMs.

\newcolumntype{P}[1]{>{\centering\arraybackslash}p{#1}}
\begin{table}[t!]
\centering
\renewcommand{\arraystretch}{2}
\captionsetup{justification=centering}
\caption{Hardware overhead of ReTern for different memory technologies}
\begin{tabular}{ |P{0.25\columnwidth}| P{0.15\columnwidth}| P{0.15\columnwidth}| P{0.15\columnwidth}|}
 \hline
  & \textbf{Energy} & \textbf{Latency} & \textbf{Area}\\
  \hline
   \textbf{SRAM} & 2.0\% & 3.2\% & $<$1\% \\
  \hline
  \textbf{ReRAM} & 2.2\% & 6.6\% & $<$1\% \\
 \hline
 \textbf{FeFET} & 2.2\% & 6.4\% & $<$1\% \\
 \hline
\end{tabular}
\label{table:overhead}
\end{table}
\section{Conclusion}
\label{sec:conclusion}

In this work, we investigate the SAF tolerance of ternary LLMs deployed on TCiM arrays. To mitigate the accuracy degradation caused by SAFs, we propose ReTern, a training-free technique that utilizes zero-fix and fault-aware sign transformation (FAST) to reduce errors due to SAFs. While zero-fix corrects SAF-induced errors in zero weights by exploiting the natural redundancy present in TCiM bitcells, FAST mitigates the SAF-induced errors in non-zero weights, thus complementing zero-fix and further curbing the accuracy degradation. We evaluate ReTern on two BitNet 1.58B ternary LLMs, with 700M and 3B parameters, deployed on TCiM hardware. Our experiments measure perplexity on the Wikitext dataset and accuracy on the PIQA and ARC (easy) benchmarks. ReTern achieves up to a 35\% reduction in perplexity and accuracy improvements of up to 4\% on PIQA and 6\% on ARC (easy). Additionally, we assess the hardware overheads of enabling ReTern at the memory macro level for 8T-SRAM, 1T-1ReRAM, and 1-FeFET TCiM arrays at the 7nm technology node. ReTern incurs mild overheads, requiring only 2\%–2.2\% additional energy, 3.2\%–6.6\% higher latency, and $<$1\% area overhead compared to the baseline. Overall, ReTern is a low overhead training-free approach to enhance the SAF tolerance of TCiM accelerators. While we have evaluated ReTern in the context of ternary LLMs, it is also applicable, in principle, to other ternary precision models such as CNNs \cite{rtn,sparse_tnn} and transformers \cite{ternary_bert} (which need to be evaluated). By targeting faults in both zero and non-zero weights, ReTern can be effective for workloads displaying a range of weight sparsity.

\section{Acknowledgements}
\label{sec:acknowledgements}

This work is supported, in part, by the Center for the Co-Design of Cognitive Systems (COCOSYS), one of seven centers in JUMP 2.0, funded by Semiconductor Research Corporation (SRC) and DARPA, Raytheon and NSF. The authors would also like to thank Doug Hyun Kim for his helpful suggestions.

\bibliography{references.bib}

\begin{thebibliography}{10}
\providecommand{\url}[1]{#1}
\csname url@samestyle\endcsname
\providecommand{\newblock}{\relax}
\providecommand{\bibinfo}[2]{#2}
\providecommand{\BIBentrySTDinterwordspacing}{\spaceskip=0pt\relax}
\providecommand{\BIBentryALTinterwordstretchfactor}{4}
\providecommand{\BIBentryALTinterwordspacing}{\spaceskip=\fontdimen2\font plus
\BIBentryALTinterwordstretchfactor\fontdimen3\font minus \fontdimen4\font\relax}
\providecommand{\BIBforeignlanguage}[2]{{%
\expandafter\ifx\csname l@#1\endcsname\relax
\typeout{** WARNING: IEEEtran.bst: No hyphenation pattern has been}%
\typeout{** loaded for the language `#1'. Using the pattern for}%
\typeout{** the default language instead.}%
\else
\language=\csname l@#1\endcsname
\fi
#2}}
\providecommand{\BIBdecl}{\relax}
\BIBdecl

\bibitem{llm_cost}
S.~Samsi \emph{et~al.}, ``From words to watts: Benchmarking the energy costs of large language model inference,'' in \emph{2023 IEEE High Performance Extreme Computing Conference (HPEC)}, 2023, pp. 1--9.

\bibitem{cim_llm1}
\BIBentryALTinterwordspacing
Y.~Wu, Z.~Wang, and W.~D. Lu, ``Pim gpt a hybrid process in memory accelerator for autoregressive transformers,'' \emph{npj Unconventional Computing}, vol.~1, no.~1, p.~4, Jul 2024. [Online]. Available: \url{https://doi.org/10.1038/s44335-024-00004-2}
\BIBentrySTDinterwordspacing

\bibitem{cim_llm2}
M.~Zhou, W.~Xu, J.~Kang, and T.~Rosing, ``Transpim: A memory-based acceleration via software-hardware co-design for transformer,'' in \emph{2022 IEEE International Symposium on High-Performance Computer Architecture (HPCA)}, 2022, pp. 1071--1085.

\bibitem{cim_llm3}
S.~Liu \emph{et~al.}, ``Hardsea: Hybrid analog-reram clustering and digital-sram in-memory computing accelerator for dynamic sparse self-attention in transformer,'' \emph{IEEE Transactions on Very Large Scale Integration (VLSI) Systems}, vol.~32, no.~2, pp. 269--282, 2024.

\bibitem{cim_llm4}
\BIBentryALTinterwordspacing
J.~B{\"u}chel \emph{et~al.}, ``Efficient scaling of large language models with mixture of experts and 3d analog in-memory computing,'' \emph{Nature Computational Science}, vol.~5, no.~1, pp. 13--26, Jan 2025. [Online]. Available: \url{https://doi.org/10.1038/s43588-024-00753-x}
\BIBentrySTDinterwordspacing

\bibitem{llm_quant}
Z.~Liu \emph{et~al.}, ``Paretoq: Scaling laws in extremely low-bit llm quantization,'' \emph{arXiv preprint arXiv:2502.02631}, 2025.

\bibitem{bitnet1}
\BIBentryALTinterwordspacing
H.~Wang, S.~Ma, L.~Dong, S.~Huang, H.~Wang, L.~Ma, F.~Yang, R.~Wang, Y.~Wu, and F.~Wei, ``Bitnet: Scaling 1-bit transformers for large language models,'' 2023. [Online]. Available: \url{https://arxiv.org/abs/2310.11453}
\BIBentrySTDinterwordspacing

\bibitem{bitnet2}
\BIBentryALTinterwordspacing
S.~Ma \emph{et~al.}, ``The era of 1-bit llms: All large language models are in 1.58 bits,'' 2024. [Online]. Available: \url{https://arxiv.org/abs/2402.17764}
\BIBentrySTDinterwordspacing

\bibitem{bitnet3}
\BIBentryALTinterwordspacing
H.~Wang, S.~Ma, and F.~Wei, ``Bitnet a4.8: 4-bit activations for 1-bit llms,'' 2024. [Online]. Available: \url{https://arxiv.org/abs/2411.04965}
\BIBentrySTDinterwordspacing

\bibitem{site_cim}
\BIBentryALTinterwordspacing
N.~Thakuria, A.~Malhotra, S.~K. Thirumala, R.~Elangovan, A.~Raghunathan, and S.~K. Gupta, ``Site cim: Signed ternary computing-in-memory for ultra-low precision deep neural networks,'' 2024. [Online]. Available: \url{https://arxiv.org/abs/2408.13617}
\BIBentrySTDinterwordspacing

\bibitem{tim_dnn}
S.~Jain, S.~K. Gupta, and A.~Raghunathan, ``Tim-dnn: Ternary in-memory accelerator for deep neural networks,'' \emph{IEEE Transactions on Very Large Scale Integration (VLSI) Systems}, vol.~28, no.~7, pp. 1567--1577, 2020.

\bibitem{ternary1}
H.~Jeong, S.~Kim, K.~Park, J.~Jung, and K.~J. Lee, ``A ternary neural network computing-in-memory processor with 16t1c bitcell architecture,'' \emph{IEEE Transactions on Circuits and Systems II: Express Briefs}, vol.~70, no.~5, pp. 1739--1743, 2023.

\bibitem{ternary2}
S.~Cheon \emph{et~al.}, ``A 2941-tops/w charge-domain 10t sram compute-in-memory for ternary neural network,'' \emph{IEEE Transactions on Circuits and Systems I: Regular Papers}, vol.~70, no.~5, pp. 2085--2097, 2023.

\bibitem{training1}
M.~Xiang \emph{et~al.}, ``Drop-connect as a fault-tolerance approach for rram-based deep neural network accelerators,'' in \emph{2024 IEEE 42nd VLSI Test Symposium (VTS)}, 2024, pp. 1--7.

\bibitem{training2}
L.~Xia, M.~Liu, X.~Ning, K.~Chakrabarty, and Y.~Wang, ``Fault-tolerant training with on-line fault detection for rram-based neural computing systems,'' in \emph{2017 54th ACM/EDAC/IEEE Design Automation Conference (DAC)}, 2017, pp. 1--6.

\bibitem{ai_index}
\BIBentryALTinterwordspacing
{Stanford University}, ``{Artificial Intelligence Index Report 2024},'' 2024, accessed: 2025-02-25. [Online]. Available: \url{https://aiindex.stanford.edu/wp-content/uploads/2024/04/HAI_2024_AI-Index-Report.pdf}
\BIBentrySTDinterwordspacing

\bibitem{mapping1}
J.~Zhang, C.~Wang, Y.~Cai, Z.~Zhu, D.~Kline, H.~Yang, and Y.~Wang, ``Wesco: Weight-encoded reliability and security co-design for in-memory computing systems,'' in \emph{2022 IEEE Computer Society Annual Symposium on VLSI (ISVLSI)}, 2022, pp. 296--301.

\bibitem{mapping2}
H.~Shin \emph{et~al.}, ``Fault-free: A framework for analysis and mitigation of stuck-at-fault on realistic reram-based dnn accelerators,'' \emph{IEEE Transactions on Computers}, vol.~72, no.~7, pp. 2011--2024, 2023.

\bibitem{mapping3}
\BIBentryALTinterwordspacing
B.~Zhang, N.~Uysal, D.~Fan, and R.~Ewetz, ``Handling stuck-at-faults in memristor crossbar arrays using matrix transformations,'' in \emph{Proceedings of the 24th Asia and South Pacific Design Automation Conference}, ser. ASPDAC '19.\hskip 1em plus 0.5em minus 0.4em\relax New York, NY, USA: Association for Computing Machinery, 2019, p. 438–443. [Online]. Available: \url{https://doi.org/10.1145/3287624.3287707}
\BIBentrySTDinterwordspacing

\bibitem{redundancy3}
\BIBentryALTinterwordspacing
B.~Li \emph{et~al.}, ``Zero-space cost fault tolerance for transformer-based language models on reram,'' 2024. [Online]. Available: \url{https://arxiv.org/abs/2401.11664}
\BIBentrySTDinterwordspacing

\bibitem{tfix}
A.~Malhotra, C.~Wang, and S.~K. Gupta, ``Tfix: Exploiting the natural redundancy of ternary neural networks for fault tolerant in-memory vector matrix multiplication,'' in \emph{2023 60th ACM/IEEE Design Automation Conference (DAC)}, 2023, pp. 1--6.

\bibitem{sparse_tnn}
A.~Marban \emph{et~al.}, ``Learning sparse \& ternary neural networks with entropy-constrained trained ternarization (ec2t),'' in \emph{2020 IEEE/CVF Conf. on Computer Vision and Pattern Recognition Workshops}, 2020.

\bibitem{fat}
S.~Zhu \emph{et~al.}, ``{FAT}: An in-memory accelerator with fast addition for ternary weight neural networks,'' \emph{{IEEE} Transactions on Computer-Aided Design of Integrated Circuits and Systems}, 2022.

\bibitem{multistate}
Q.~Cao \emph{et~al.}, ``\BIBforeignlanguage{en}{Nonvolatile multistates memories for high-density data storage},'' \emph{\BIBforeignlanguage{en}{ACS Appl. Mater. Interfaces}}, vol.~12, no.~38, pp. 42\,449--42\,471, Sep. 2020.

\bibitem{redundancy1}
L.~Xia \emph{et~al.}, ``Stuck-at fault tolerance in rram computing systems,'' \emph{IEEE Journal on Emerging and Selected Topics in Circuits and Systems}, vol.~8, no.~1, pp. 102--115, 2018.

\bibitem{redundancy2}
\BIBentryALTinterwordspacing
C.~Kim, D.~Yoon, T.~Kim, Y.~Jeong, K.~Kim, K.~Koh, and E.~Pak, ``Analog computing for {AI} sometimes needs correction by digital computing: Why and when,'' in \emph{NeurIPS 2024 Workshop Machine Learning with new Compute Paradigms}, 2024. [Online]. Available: \url{https://openreview.net/forum?id=GebgnS1TdT}
\BIBentrySTDinterwordspacing

\bibitem{rtn}
\BIBentryALTinterwordspacing
Y.~Li \emph{et~al.}, ``Rtn: Reparameterized ternary network,'' \emph{Proceedings of the AAAI Conference on Artificial Intelligence}, vol.~34, no.~04, pp. 4780--4787, Apr. 2020. [Online]. Available: \url{https://ojs.aaai.org/index.php/AAAI/article/view/5912}
\BIBentrySTDinterwordspacing

\bibitem{fault_testing}
C.-Y. Chen, H.-C. Shih, C.-W. Wu, C.-H. Lin, P.-F. Chiu, S.-S. Sheu, and F.~T. Chen, ``Rram defect modeling and failure analysis based on march test and a novel squeeze-search scheme,'' \emph{IEEE Transactions on Computers}, vol.~64, no.~1, pp. 180--190, 2015.

\bibitem{700m_model}
1bitLLM, ``Bitnet b1.58-large,'' \url{https://huggingface.co/1bitLLM/bitnet_b1_58-large}, accessed: 2025-03-07.

\bibitem{3b_model}
------, ``Bitnet b1.58-3b,'' \url{https://huggingface.co/1bitLLM/bitnet_b1_58-3B}, accessed: 2025-03-07.

\bibitem{piqa}
\BIBentryALTinterwordspacing
Y.~Bisk, R.~Zellers, R.~L. Bras, J.~Gao, and Y.~Choi, ``Piqa: Reasoning about physical commonsense in natural language,'' 2019. [Online]. Available: \url{https://arxiv.org/abs/1911.11641}
\BIBentrySTDinterwordspacing

\bibitem{arc}
\BIBentryALTinterwordspacing
P.~Clark \emph{et~al.}, ``Think you have solved question answering? try arc, the ai2 reasoning challenge,'' 2018. [Online]. Available: \url{https://arxiv.org/abs/1803.05457}
\BIBentrySTDinterwordspacing

\bibitem{heterogenous}
S.~Jain \emph{et~al.}, ``A heterogeneous and programmable compute-in-memory accelerator architecture for analog-ai using dense 2-d mesh,'' \emph{IEEE Transactions on Very Large Scale Integration (VLSI) Systems}, vol.~31, no.~1, pp. 114--127, 2023.

\bibitem{ptm}
\BIBentryALTinterwordspacing
Accessed: August 2024. [Online]. Available: \url{https://asap.asu.edu/}
\BIBentrySTDinterwordspacing

\bibitem{rram}
Z.~Jiang \emph{et~al.}, ``A compact model for metal–oxide resistive random access memory with experiment verification,'' \emph{IEEE Transactions on Electron Devices}, vol.~63, no.~5, pp. 1884--1892, 2016.

\bibitem{fefet}
A.~K. Saha and S.~K. Gupta, ``Modeling and comparative analysis of hysteretic ferroelectric and anti-ferroelectric fets,'' in \emph{2018 76th Device Research Conference (DRC)}, 2018, pp. 1--2.

\bibitem{fefet_experiments}
K.~Ni \emph{et~al.}, ``In-memory computing primitive for sensor data fusion in 28 nm hkmg fefet technology,'' in \emph{2018 IEEE International Electron Devices Meeting (IEDM)}, 2018, pp. 16.1.1--16.1.4.

\bibitem{fefet_sims}
A.~K. Saha \emph{et~al.}, ``Ferroelectric thickness dependent domain interactions in fefets for memory and logic: A phase-field model based analysis,'' in \emph{2020 IEEE International Electron Devices Meeting (IEDM)}, 2020, pp. 4.3.1--4.3.4.

\bibitem{bitcells}
\BIBentryALTinterwordspacing
C.~Wang, J.~Victor, and S.~K. Gupta, ``Comparative evaluation of memory technologies for synaptic crossbar arrays -- part i: Robustness-driven device-circuit co-design and system implications,'' 2024. [Online]. Available: \url{https://arxiv.org/abs/2307.04261}
\BIBentrySTDinterwordspacing

\bibitem{para_res}
X.~Chen \emph{et~al.}, ``Modeling and circuit analysis of interconnects with tas2 barrier/liner,'' in \emph{2021 Device Research Conference (DRC)}, 2021, pp. 1--2.

\bibitem{para_cap}
J.~H.-C. Chen \emph{et~al.}, ``Interconnect performance and scaling strategy at 7 nm node,'' in \emph{IEEE International Interconnect Technology Conference}, 2014, pp. 93--96.

\bibitem{neurosim}
X.~Peng, S.~Huang, Y.~Luo, X.~Sun, and S.~Yu, ``Dnn+neurosim: An end-to-end benchmarking framework for compute-in-memory accelerators with versatile device technologies,'' in \emph{2019 IEEE International Electron Devices Meeting (IEDM)}, 2019, pp. 32.5.1--32.5.4.

\bibitem{ternary_bert}
\BIBentryALTinterwordspacing
W.~Zhang, L.~Hou, Y.~Yin, L.~Shang, X.~Chen, X.~Jiang, and Q.~Liu, ``{T}ernary{BERT}: Distillation-aware ultra-low bit {BERT},'' in \emph{Proceedings of the 2020 Conference on Empirical Methods in Natural Language Processing (EMNLP)}, B.~Webber, T.~Cohn, Y.~He, and Y.~Liu, Eds.\hskip 1em plus 0.5em minus 0.4em\relax Online: Association for Computational Linguistics, Nov. 2020, pp. 509--521. [Online]. Available: \url{https://aclanthology.org/2020.emnlp-main.37/}
\BIBentrySTDinterwordspacing

\end{thebibliography}
\bibliographystyle{IEEEtran}

\end{document}